\RequirePackage{ifpdf}
\documentclass{PoS}
\usepackage[authoryear,square]{natbib}
\bibpunct{(}{)}{;}{a}{}{,}

\title{The ionised, radical and molecular Milky Way: spectroscopic surveys with the SKA}

\ShortTitle{The ionised,radical and molecular Milky Way}

\author{
\speaker{Mark Thompson}$^1$, 
Henrik Beuther$^2$,
Clive Dickinson$^{3}$,
Joseph C. Mottram$^{5}$
Pamela Klaassen$^{5,6}$
Adam Ginsburg$^{7}$
Steve Longmore$^{4}$
Anthony Remijan$^{8}$,
Karl Menten$^{9}$\\ 
$^1$University of Hertfordshire; 
$^2$MPIA Heidelberg;
$^3$University of Manchester;
$^4$Liverpool John Moores University;
$^5$Leiden University;
$^6$UK Astronomy Technology Centre;
$^7$European Southern Observatory;
$^8$National Radio Astronomy Observatory;
$^9$MPIfR Bonn;
\\
E-mail: \email{m.a.thompson at herts.ac.uk}
}

\abstract{The bandwith, sensitivity and sheer survey speed of the SKA offers unique potential for deep spectroscopic surveys of the Milky Way.  Within the frequency bands available to the SKA lie many transitions that trace the ionised, radical and molecular components of the interstellar medium and which will revolutionise our understanding of many physical processes. In this chapter we describe the impact on our understanding of the Milky Way that can be achieved by spectroscopic SKA surveys, including ``out of the box'' early science with radio recombination lines, Phase 1 surveys of the molecular ISM using anomalous formaldehyde absorption, and full SKA surveys of ammonia inversion lines.}

\FullConference{
Advancing Astrophysics with the Square Kilometre Array\\
June 8-13, 2014\\
Giardini Naxos, Sicily, Italy}

\newcommand{\skipthis}[1]{}

\newcommand{\arcsec}{$^{\prime\prime}$}
\newcommand{\arcmin}{$^{\prime}$}
\newcommand{\degr}{$^{\circ}$}

\begin{document}

\section{Introduction}

Over the last decade there has been a renaissance in multiwavelength surveys of the Milky Way, exploiting new facilities and instrumentation to conduct wide area surveys that are over an order of magnitude deeper and at much higher angular resolution than their predecessors. The combination of this wealth of survey data is beginning to revolutionise our understanding of the complex cycle that relates the interstellar medium (ISM) to star formation. Two factors play a  part in the process: firstly   the different wavelengths covered by each survey trace (very) different components of the ISM; and secondly the surveys have essentially close to matching angular resolution ($\sim$10--20\arcsec) over the bulk of their combined wavelength range.

%These surveys cover over three decades of wavelength from the near-infrared (UKIDSS/VVV); through the mid-infrared (GLIMPSE/GLIMPSE360/MIPSGAL; Benjamin et al 2003; Whitney et al 2011; Carey et al 2009) to the far-infrared (\emph{Herschel} Hi-GAL; Molinari et al 2010a), sub-mm (ATLASGAL; Schuller et al 2009) and radio (CORNISH/CORNISH-S; Hoare et al 2012).

Spectroscopic surveys are a key piece of the puzzle, as they trace the detailed kinematics of the ISM and reveal the complex 3-dimensional structure of the Milky Way, providing the crucial third dimension to continuum surveys. The extra complexity and depth required by spectroscopic surveys means that they are in general one ``generation'' behind the most recent corresponding continuum surveys. For example, the current state of the art in surveys of the atomic and molecular components of the ISM are the $\sim$ 1\arcmin\ resolution surveys of HI and CO (e.g.~the International Galactic Plane Survey, the Galactic Ring Survey \& the FCRAO Outer Galaxy Survey). By the time of the SKA these projects are likely to be superseded by higher resolution and more sensitive HI/CO surveys (with JVLA, ASKAP, MeerKAT, JCMT and possibly CCAT) that will match the resolution of the current far-infrared-millimetre wave continuum surveys. 

However, there are a number of areas that will \emph{not} be addressed by forthcoming spectroscopic surveys, where the sheer potential of the SKA in mapping faint radio-wavelength lines will play an important role across all the components of the ISM. As one of the principal raisons d'\^etre of the SKA is radio spectroscopy, the compact $\sim$1 km cores of SKA1-MID and SKA1-SUR (leading to the $\sim$ 4 km core of the full SKA) are optimised for brightness temperature sensitivity. Combined with the relatively large FOV of the 15m dishes, this makes the SKA an incredibly powerful facility for wide area spectroscopic surveys of not just HI but all radio-wavelength lines. Within the wavelength ranges available in SKA Bands 1--5 there are many lines that trace multiple components of the ISM, including thousands of radio recombination lines (RRLs), several lines from light hydride radicals (OH and CH), the $^{3}$He$^{+}$ hyperfine line and the two anomalous absorption lines of o-H$_{2}$CO. These faint radio phenomena which were all discovered in the 1960s can be deployed as standard tools in the SKA era to study the ionised, radical and molecular components of the ISM in unprecedented detail.

In this chapter we will describe the ISM science that can be achieved with the SKA by mapping (classical) RRLs, hydride radical lines and anomalous absorption formaldehyde lines. These studies have tremendous scope for improving our understanding of a range of processes within the Milky Way, from accretion processes in massive star formation, stellar feedback into the ISM and the origin of the warm diffuse ionised ISM. We refer readers interested in  HI, diffuse RRLs and OH masers to the chapters by \citet{mcclure-griffiths2014}, \citet{oonk2014} and \citet{etoka2014} respectively. We will outline a series of strawman projects that could be accomplished by the SKA in ``early science'' mode, showing how the expanding capabilities of SKA1-MID and SKA1-SUR naturally lead to enhanced science capabilities. Finally we briefly dwell upon the potential that a fully frequency-capable SKA ($\nu\le$ 24 GHz) has for large area surveys of NH$_{3}$.

%In this chapter we will show that the SKA is an ideal survey machine for both wide-area RRL studies of the diffuse extended  ISM and high angular resolution observations of the dense gas found within young HII regions. These studies have tremendous scope for improving our understanding of a range of processes within the Milky Way, from accretion processes in massive star formation, stellar feedback into the ISM and the origin of the warm diffuse ionised ISM. Combined with SKA HI studies  and ALMA/CCAT molecular line observations, SKA RRL surveys will finally result in a complete picture of the interplay between ionised, atomic and molecular phases of the ISM.

\section{Radio Recombination lines: the kinematics of the ionised ISM}

Radio recombination lines (RRLs) arise from atomic transitions between large principal quantum numbers (typically $n\ge40$), where the small difference between energy levels means that the emitted photons are of radio wavelength.  The usual convention to describe the transition  $n+m\,\rightarrow\,n$ is $n\alpha$, $n\beta$, $n\gamma$\ldots for $m = 1, 2, 3, \ldots$, hence $\alpha$   transitions (which are the most probable) correspond to a change in the quantum number of one. Despite the intrinsic faintness of their high quantum number transitions  RRLs are one of the best tracers of astrophysical plasmas due to their well-understood physics \citep{gordon2002}, their immunity to extinction (unlike the Balmer H$\alpha$ or OIII fine structure lines) and the sheer line density of their spectra (many lines of H, He and C are found close together in frequency). From RRL measurements one can determine many physical properties of the ionised gas, e.g. temperature and  electron density \citep{gordon2002,brocklehurst1972}, metallicity \citep{balser2001}, the hardness of the illuminating UV spectrum \citep{roshi2012}, and potentially magnetic field strengths from C RRL linewidths \citep{roshi2007}. By careful selection of various lines it is also possible to use RRLs to trace the kinematics of ionised gas from the extended low density medium to the densest parts of young compact HII regions and planetary nebulae. 

The main limitation of current RRL studies is the tradeoff between brightness temperature sensitivity and angular resolution. Single dish studies like SIGGMA on Arecibo \citep{liu2013} and HIPASS on Parkes \citep{alves2012} offer mK sensitivity and trace electron densities down to $n_{e}\sim$ 10\,cm$^{-3}$ but with an angular resolution of several arcminutes. Interferometric surveys like the JVLA THOR survey \citep{bihr2013}, which is mapping a 52\degr\ long strip of the  northern Galactic Plane, reach angular resolutions of 10\arcsec\, but are only sensitive to $\sim$ 1 K lines which corresponds to $n_{e} \stackrel{>}{\sim} 1000$\,cm$^{-3}$ (and moreover, only for the brightest n$\alpha$ lines). Obviously, single-dish apertures such as FAST are more suited to trace the truly extended diffuse medium, but the SKA has the unique potential for high resolution studies of the intermediate density gas ($n_{e} \sim 100$ cm$^{-3}$) at the interface between  HII regions and the diffuse ionised interstellar medium.  This would enable the morphology and kinematics of the HII region boundaries to be explored at comparable resolution to HI \& CO observations, plus determine the spectrum of the escaping radiation via simultaneous detections of H and He RRLs.

The sheer sensitivity, broadband feeds and highly capable correlator  make the SKA an ideal instrument to observe RRLs. The widths of the lines are a few 10s of km\,s$^{-1}$, so they can be comfortably observed in the continuum mode of the SKA correlator. Coupled with the wide bandwidths and density of RRL spectra, this means that there are literally thousands of RRLs available to be observed in SKA1-MID and SUR Bands 1, 2 \& 3 --- falling to hundreds of lines in SKA1-MID Band 5. Secondly, for  detection experiments the well understood physics governing the line frequencies (essentially the Rydberg equation) implies that the lines can easily be stacked to improve the signal to noise. As the line separations are fixed, adaptive stacking at different input V$_{\rm lsr}$ can be used to retrieve line detections even if the line of sight velocity of the gas is not initially known. The larger instantaneous bandwidths of SKA1-MID over SKA1-SUR make SKA1-MID competitive in mapping speed once line stacking is taken into consideration. Indeed, assuming that half the lines in the passband are free from RFI, a line-to-continuum ratio of 0.02 and channel width of 10 km\,s$^{-1}$, SKA1-SUR and SKA1-MID have a stacked RRL spectroscopic survey speed for H$n\alpha$ lines that is within a factor 2 of the VLA (not JVLA!) \emph{continuum} survey speed. It will be well within the capability of the Phase 1 SKA to perform spectroscopic RRL surveys that have the same sensitivity to ionised gas that present day surveys, e.g.~the VLA Galactic Plane Survey \citep{stil2006}, achieve in continuum.

Thirdly, the broadband feeds imply the simultaneous detection of multiple $n$ and $m$ lines from the same atomic species. This enables detailed radiative transfer models to be created that constrain the effects of stimulated emission and  departures from LTE. In addition, multiple lines at different $n$ values can be used to estimate the pressure broadening effects and determine the velocity and density structure of the ionised gas.  Finally, the SKA will also observe  RRLs from multiple atomic species, principally Helium and Carbon, as these lines are found within 2 MHz of corresponding H RRLs. Helium lines are typically $\sim$8 percent of the brightness of Hydrogen RRLs, but are valuable tracers of both the metallicity of the ionised gas \citep[e.g.][]{balser2001} and the hardness of the illuminating UV spectrum \citep{roshi2012}. More than 90\% of the observed  $^{4}$He in the ISM of the Milky Way was produced via primordial nucleosynthesis \citep{wilson1994}. Observations of He RRLs at high angular resolution are a unique window into this process,  allowing the excitation and total abundance to be properly modelled. Carbon RRLs are produced in the PDRs surrounding HII regions and so these lines are a valuable tracer of the physical conditions in this gas, particularly when combined with observations of the 158 $\mu$m CII line \citep{natta1994}. Their line brightnesses are typically $\sim$30\% of H RRLs. The non-thermal linewidths of C RRLs may also allow magnetic field strengths to be measured \citep{roshi2007}.

\section{Anomalous formaldehyde absorption: tracing the volume density of H$_{2}$ via silhouettes on the CMB}

The lowest two transitions of ortho formaldehyde (at 4.8 and 14.4 GHz) have a curious property, in that collisional excitation of these transitions drives an ``anti-inversion''  which cools the lines with respect to the CMB temperature. The transitions can then absorb CMB photons and appear in absorption against the CMB. This phenomenon is known as anomalous absorption and was first observed by \citet{palmer1969} and theoretically explained by \citet{townes1969}. Anomalous formaldehyde absorption is an incredibly useful tool for studying molecular clouds. As the illuminating source is isotropic and fills the Universe, anomalous absorption is distance-independent and wholly dependent on the number of absorbing molecules. Moreover, the ratio of the 4.8 and 14.4 GHz lines is highly sensitive to the H$_{2}$ \emph{volume density} \citep{mangum2008,ginsburg2011}, making the combination of these transitions an effective molecular densitometer analogous with the molecular thermometer of the 218 GHz para formaldehyde K-doublet.  This densitometer is unaffected by the sub-thermal excitation, line trapping or optical depth effects that plague CO observations, allowing the H$_{2}$ mass of the absorbing source to be determined within 0.3 dex. So with sufficient sensitivity and angular resolution (to couple the synthesised beam to the absorbing source) one can use anomalous absorption to accurately measure the density and mass of molecular gas clouds from the Milky Way \citep{ginsburg2011} to  local galaxies \citep{mangum2008} and beyond to the  high-redshift Universe \citep{zeiger2010,darling2012}.

However, these absorption lines are faint and narrow, and thus require long integration times to adequately detect. This means that they cannot currently be used in wide area surveys --- unlike the bright CO rotational transitions which have been widely used to map molecular emission in the Milky Way. For example, to detect the absorption lines from Galactic clouds requires a sensitivity of  $\sim$0.1 K, which takes at least a full 12-hour track with the VLA \citep{evans1987}. But with  SKA1-MID it will become possible to detect these lines in only 1--2 hours (and minutes with the SKA) making it feasible to survey the Milky Way's molecular clouds at much better angular resolution than existing single-dish CO surveys, and with none  of the excitation or optical depth issues that affect CO surveys. In addition to anomalous absorption, a wide area SKA survey will also observe ``non-anomalous'' absorption against hundreds to thousands of bright continuum sources within and without the Milky Way (e.g. radio galaxies and HII regions). Many of the compact HII regions will have accurately measured trigonometric parallaxes from associated maser sources \citep[see the chapter by][]{green2014}. This offers the potential to conduct 3D tomography of the molecular ISM by using a network of hundreds of known illuminating sources with well-characterised heliocentric distances.

It must be stressed that the current SKA baseline design for SKA1-MID Band 5 does not cover the 14.4 GHz H$_{2}$CO line, but a  modest extension  to the upper frequency limit of Band 5 from 13.8 GHz to 14.4 GHz would enable the line ratios to be measured  and \emph{uniquely} permit  molecular gas volume densities to be determined across the Milky Way and beyond.

\section{Hydride radicals: thermal OH and CH}

Here we discuss thermal emission from the two main hydride radical species in the SKA bands: OH at 1.7 GHz in SKA1-MID band 2 (band 3 in SKA1-SUR) and the 0.7 \& 3.3 GHz CH lines in SKA1-MID bands 1 and 4 (band 3 in SKA1-SUR). There are a number of current OH surveys, the single-dish SPLASH in the Southern Hemisphere \citep{dawson2014} and the interferometric THOR survey in the Northern Hemisphere \citep{bihr2013}, with a planned deeper ASKAP survey (GASKAP: \citealt{dickey2013}).  Thermal emission from OH is often overlooked in favour of the much brighter non-thermal OH masers \citep[see the Chapter by][]{etoka2014} --- nevertheless as OH has a largely constant abundance across diffuse and translucent molecular clouds, has four closely spaced transitions, and is thermalised even at low densities ($n_{\rm crit}\le 4$ cm$^{-3}$) this molecule is a useful tracer of  the temperature and density of the neutral ISM. Thermal OH may even allow the detection of CO-dark gas hinted at by gamma ray emission \citep[e.g.][]{abdo2010}, \emph{Herschel} CII observations \citep{langer2014} and single-dish OH pencil beam studies \citep{allen2012}. Again, the SKA will bring the benefit of its brightness temperature sensitivity to the study of thermal OH, reaching and order of magnitude greater sensitivity and spatial resolution over GASKAP.

CH was one of the first molecular radicals detected in the interstellar medium via optical absorption spectroscopy, and is a very good tracer of H$_{2}$ column density in UV-dominated regions \citep{sheffer2008}  and the diffuse ISM \citep{Qin2010}. The 0.7 and 3.3 GHz lines available to the SKA have also been postulated to be a sensitive probe of changes in fundamental constants, although as these lines are subject to non-LTE effects caution must be taken to model the lines carefully, possibly also using higher frequency data from \emph{Herschel} or SOFIA. The SKA enables wide area and sensitive surveys of CH, and in combination with OH and H$_{2}$CO anomalous absorption allows the entire dynamic range of the molecular ISM to be traced from PDRs, to diffuse and dense H$_{2}$ clouds.

\section{Potential SKA spectroscopic surveys}

In this section we describe potential RRL survey projects that the SKA could carry out, paying particular attention to the science outcomes that are enabled by the different components of the SKA and the possibilities for Early Science. 

%The overall strategy that maximises scientific return is a ``wedding cake'' of surveys, with large area low frequency observations targeting the low density ionised ISM and progressively higher frequency and higher angular resolution observations of high density regions. 
In the following we  concentrate on SKA1-MID and -SUR Band 2 and SKA1-MID Band 5, as these  offer the greatest potential for studies of the diffuse and dense ionised ISM, hydrides and anomalous formaldehyde absorption,  but as mentioned earlier there are many RRLs present in the other SKA bands that could be part of commensal studies. In the following calculations we have assumed the noise and imaging performance values for the Baseline Design of SKA1-MID and SKA1-SUR \citep{dewdney2013} given in the SKA1 Imaging Science Performance memo \citep{braun2014}. We have also conservatively  assumed that in SKA1-MID and -SUR Band 2 we will be able to stack 25 RRLs using the 1 GHz bandwidth of SKA1-MID and 12 RRLs using the 500 MHz bandwidth of SKA1-SUR. Thus, in a 1 hour integration SKA1-MID is able to reach an rms flux of 106 $\mu$Jy per 10 km\,s$^{-1}$ channel, which translates into a stacked RRL sensitivity for $\alpha$ lines of 21 $\mu$Jy per 10 km\,s$^{-1}$ channel. Similarly SKA1-SUR is able to achieve rms fluxes of 430 $\mu$Jy and 124 $\mu$Jy per 10 km\,s$^{-1}$ channel in unstacked and stacked data respectively. Comparing these values to the relative fields-of-view of SKA1-MID and SKA1-SUR, it can be seen that SKA1-MID is highly  competitive for RRL mapping over small areas in Band 2 due to its larger instantaneous bandwidth and low SEFD, although SKA1-SUR has a faster mapping speed over areas larger than a single PAF tile.

\subsection{Early Science: the structure and kinematics of the most luminous HII regions and their impact on the ISM}

The most luminous HII region complexes in the Milky Way have  been identified using a combination of WMAP, Spitzer and MSX data by \citet{murray2010}. The 18 most luminous of these complexes are responsible for just over half the total Galactic ionising flux, and it is thought to be the UV photons ``leaking'' from the HII region complexes that are responsible for the diffuse warm ionised medium \citep[e.g.~][]{liu2013}. This hypothesis is supported by wide area RRL mapping at $\sim$15\arcmin\ resolution \citep{alves2012} which finds that the distribution of the diffuse medium is correlated with HII regions. However, constraints on the ionising spectrum from He RRLs are not fully consistent with this hypothesis \citep{roshi2012}. Thus higher resolution observations of H RRLs and more sensitive observations to detect He RRLs are required.  A better theoretical understanding of the interplay between HII region boundaries and the diffuse surrounding medium would also complement further observations, particularly at the smaller physical scales available to the SKA.

The SKA has the unique potential for deep, high resolution studies of the intermediate density gas ($n_{e} \sim 100$ cm$^{-3}$) at the interface between  HII regions and the diffuse medium.  This would enable the morphology and kinematics of the HII region boundaries to be explored at comparable resolution to HI \& CO observations, plus determine the spectrum of the escaping radiation via simultaneous detections of H and He RRLs. Additionally, as the RRLs in SKA1-MID and -SUR Band 2 do not become appreciably pressure broadened until electron densities reach $n_{e} \simeq 1400$ cm$^{-3}$, it will also be possible to produce velocity-resolved maps of the electron density (strictly $n_{e}^{2}$) and temperature of the ionised gas within all but the densest parts of the giant HII complexes. A detailed picture of the density distribution and kinematics of HII regions  is needed to understand their evolution --- in particular the relative role of radiation pressure and stellar winds, which  has serious implications for the interpretation of galaxy population synthesis models \citep{verdolini2013}.

To detect H RRL emission from gas  with  $n_{e}$ a few 100 cm$^{-3}$ requires  a brightness temperature sensitivity of  $\sim$ 0.1 K,  integrating over a 30 km\,s$^{-1}$ line and assuming a 1 pc column \citep{alves2012}. With Early Science SEFDs of  14.2 Jy for SKA1-SUR and 3.4 Jy for SKA1-MID it is possible to achieve  this sensitivity over a 30\arcsec\ beam on the order of 50 hours and 4 hours integration time respectively. By stacking the He$n\alpha$ and C$n\alpha$ lines it will be possible to achieve a detection to the same (or slightly deeper) gas densities as the (unstacked) Hydrogen lines. By further stacking the Hydrogen lines it is possible to reach densities as low as 50 cm$^{-3}$ in the same integration times. These observations will reveal the full gamut of $\alpha$, $\beta$, $\gamma$ lines at higher densities, probing deeper into the HII regions and allowing finer velocity and spatial resolution (as the RRL brightness temperatures go as $n_{e}^{2}$) to study the distribution and kinematics of the ionised gas on smaller spatial and velocity scales.

A project to map most of the Murray \& Rahman complexes is eminently feasible in SKA Early Science, taking of the order of 200 hours to map  the top 10 most luminous complexes (small complexes with MID and larger ones with SUR). These observations can  be done ``out of the box'' at the workhorse SKA1-MID and -SUR Band 2 frequencies  without the  need to commission spectral zoom modes, effectively demonstrating the potential of the SKA for deeper and wider line surveys.

\subsection{SKA Phase 1: Galactic Plane surveys in Bands 2 and 5}

With  Phase 1 capability it becomes feasible to extend the Early Science Band 2 studies into a much deeper full Galactic Plane Survey for recombination lines. A Band 2 RRL and thermal OH survey of the  Galactic Plane can be carried out commensally with the HI survey described in the chapter by \citet{mcclure-griffiths2014}  --- rendering simultaneous maps of the atomic, ionised and molecular radical ISM in breathtaking detail. If we assume a survey of adjoining 4\degr\ wide SKA1-SUR tiles along the Galactic Plane  and a nominal dwell time of 50 hours per tile \citep[see][]{mcclure-griffiths2014} then a survey of the full Galactic Plane visible to SKA would take on the order of $\sim$ 2000 hours. With 50 hours per tile and an SEFD of 7.1 Jy for SKA1-SUR it would be possible to reach  electron densities of  $\sim$70 cm$^{-3}$ for H lines without line stacking and 40 cm$^{-3}$ with line stacking.

It is worth noting that the  stacked RRL spectral datacube would allow H RRL emission to be detected to twice as deep a continuum depth as the 1.4 GHz International Galactic Plane Survey (assuming a line-to-continuum ratio of 0.02). Such a survey would allow the creation of a detailed position-velocity map of the ionised ISM, separating the synchrotron contribution from that of the free-free emission. The RRL Galactic Plane survey will result in two main products: a continuum  map from line-free channels (which will have approximately 1.5 $\mu$Jy rms, assuming a 300 MHz continuum bandwidth) and RRL spectral datacubes for individual and stacked lines. By combining these two maps the synchrotron contribution to the continuum can be determined \citep[e.g.][]{alves2012}, resulting in a position-velocity map of the  free-free emission that matches the resolution of atomic, molecular and dust surveys. In addition to HII regions, many Planetary Nebulae (PNe) would be revealed.  PNe have been mainly overlooked in existing RRL surveys due to the limited angular resolution and resulting beam dilution. RRL observations offer a novel means of determining the distance to the Planetary Nebulae via the determination of their expansion rate \citep[e.g.~][]{gomez1989,gulyaev2003}. Multiple RRLs observed simultaneously will be the key to disentangling pressure broadening effects from the observed lines, which is one of the reasons why this approach has not been more commonly used.

The main aim of a Band 5 SKA1-MID survey would be to map the molecular ISM of the Milky Way using anomalous formaldehyde absorption but, as in the  Band 2 survey described above, significant additional benefits can also be obtained simultaneously. The 2$\times$2.5 GHz bandwidth in Band 5 allows the simultaneous detection of  75 H$n\alpha$, He$n\alpha$ and C$n\alpha$ RRLs. These lines are not appreciably pressure broadened until $n_{e}\sim 10^{5}$ cm$^{-3}$ and so they can be used to study the kinematics of compact HII regions and young dense PNe.  As yet unexplained highly turbulent motions are observed towards a number of compact and ultracompact HII regions \citep{keto2008}, which may be due to  trapping  of ionised accretion flows \citep{galvan-madrid2011} or ionised outflows \citep{klaassen2013}. The $^{3}$He$^{+}$ hyperfine line also lies within Band 5 at a rest frequency of 8.7 GHz. A large survey of HII regions and PNe in this line would enable constraints to be placed on the primordial abundance of $^{3}$He via Big Bang nucleosynthesis \citep{bania2007} and perhaps provide a solution to the ``$^{3}$He problem'' \citep{guzman-ramirez2013}. Moreover, a survey to the depth required to detect anomalous absorption would also result in a very deep 5 GHz continuum survey ($\sim$ 0.4 $\mu$Jy), which would revolutionise the study of radio stars in the Milky Way \citep[see the chapter by][]{umana2014}.

The necessary sensitivity for an anomalous absorption survey is $\sim$0.1--0.2 K rms over a channel width of 0.1 km\,s$^{-1}$ (required to detect the narrow H$_{2}$CO lines. With SKA1-MID it is possible to reach this sensitivity over a 15\arcsec\ beam in on the order of 1 hour integration time, which implies a roughly 100 deg$^{2}$ survey could be achieved in 1000 hours at 4.8 GHz. To enable the H$_{2}$ volume density to be determined requires further observations of the 14.4 GHz line and the most efficient way to achieve these would be targeted followups of regions where 4.8 GHz absorption is observed. These observations would also map out the high frequency RRL emission around these regions. These followups would take on the order of 500 hours to complete, leading to a total survey time of $\sim$1500 hours. Without the followup of the 14.4 GHz H$_{2}$CO line, the 4.8 GHz observations can only constrain the H$_{2}$CO column density rather than the H$_{2}$ volume density, hence the extension of  SKA1-MID Band 5  to higher frequency is crucial to add this important capability. This project would lead to the most comprehensive map of the molecular gas in the southern Milky Way to date, with uniform sensitivity to gas from volume densities $n(\rm{H}_{2})$ of 10$^{2.5}$--10$^{6}$ cm$^{-3}$ and accuracy of $\sim$0.3 dex.

\subsection{The full SKA}

With the deployment of the full SKA it will become possible to conduct
wide-area maps of RRLs from ionised gas of below 50 cm$^{-3}$, routinely
detecting transitions from $n\alpha$, $\beta$ and $\gamma$ transitions from
H, He and C. The deployment of frequency bands up to 25 GHz will permit
lines up to H64$\alpha$ to be observed, which extend the gas densities that
can be traced up to 10$^{6}$  cm$^{-3}$ and allow the study of the bulk of
ionised jets and winds from low mass and high mass stars \citep{hoare2004}.
There is particular synergy here with ALMA studies of higher frequency
millimetre-wave recombination lines and SKA2 observations of 10--25 GHz
lines. Millimetre wave RRLs are more subject to non-LTE effects than
microwave lines and the comparison of their velocity resolved spectra can
illustrate inflow and outflow within optically thick ultra and
hyper-compact HII regions \citep{peters2012}. Only the SKA will have the
surface brightness sensitivity to match ALMA high resolution observations.

In addition, the 24 GHz inversion transitions of ammonia (including both
$^{14}$NH$_{3}$ and $^{15}$NH$_{3}$ isotopologues) will also become
available. Using the antennas from the 4 km compact core would permit a
deep Galactic Plane survey of ammonia to be carried out at $\sim$2\arcsec\
resolution, i.e. comparable to ALMA Bands 6/7 in compact configuration.
Ammonia is a molecular thermometer \emph{par excellence}, and there are
tremendous synergies between ALMA molecular observations and the accurate
kinetic temperatures that would result from wide area SKA surveys. The
sensitivity of the full SKA implies that to detect NH$_{3}$ lines of  a few
0.1 K brightness temperature would take on the order of five minutes per
pointing. Although the primary beam at 24 GHz is only 0.08 deg$^{2}$, such
a short integration time implies a survey speed of $\sim$ 1
deg$^{2}$\,hr$^{-1}$, making it feasible to carry out a 2\arcsec\
resolution survey of the Galactic Plane in only a few hundred hours.

\setlength{\bibsep}{0.0pt}
\bibliographystyle{apj}
\bibliography{thompson_ska_chapter}
\end{document}